# New third-family flavor physics: Vertex corrections

B. HOLDOM[1]

*Department of Physics*
*University of Toronto*
*Toronto, Ontario*
CANADA  M5S 1A7

ABSTRACT

As described previously, a new massive gauge boson ($X$) coupling only to the third family produces a tantalizing pattern of deviations away from the standard model. These include increasing $\Gamma_b/\Gamma_h$ and decreasing the $\alpha_s(M_Z)$ extracted from $\Gamma_h/\Gamma_\ell$. We review the status of these $X$-$Z$ mixing effects. We then calculate $X$ boson induced vertex corrections to $Z$ partial decay widths and to $t\bar{t}$ production in $p\bar{p}$ colliders.

---

[1] holdom@utcc.utoronto.ca

The third family, and its place in the standard model, is not as well tested as the first and second families. In fact there is a growing realization that the precision electroweak experiments still a leave a large opening for new flavor physics associated with the third family. There are also reasons for believing that the lightest new nonuniversal gauge interactions, should they exist, will couple preferentially to the heaviest family. Some model independent analyses have appeared [1] and these are useful in elucidating the full range of possible effects. It is also useful to have definite theoretical targets to aim at when trying to understand the significance of experimental results. (For other examples of targets see [2].) This note will explore further a minimal extension of the standard model to include a nonuniversal flavor interaction [3]. A massive $U(1)$ gauge boson denoted by $X$ is introduced which couples to the third family but not to lighter families. We do not consider here the effects of the small fermion mass mixing between families; for some discussion see [3] and [4].

The $X$ boson we consider couples to the following current.

$$J_\mu^X \;=\; \bar{t}\gamma_\mu\gamma_5 t \;+\; \bar{b}\gamma_\mu\gamma_5 b \;+\; \bar{\tau}\gamma_\mu\tau \;+\; \bar{\nu}_{\tau L}\gamma_\mu\nu_{\tau L} \tag{1}$$

The existence of an $X$ boson was noted first in a dynamical model for the top mass [5], before its implications for precision experiments were considered.[2] The fact that the $X$ boson has axial couplings to quarks is related to the manner in which the $X$ receives a mass. It also leads to $X$-$Z$ mass mixing generated by a top quark loop, and this is what causes a shift of the $Z$ couplings to the third family [3]. The fact that the $X$ boson has vector rather than axial couplings to the $\tau$ is dictated by the observed universality in the $Z$ partial widths to $e$, $\mu$, and $\tau$. The latter constrains the $Z\tau\bar\tau$ axial coupling much more than the vector coupling since $\delta\Gamma \propto g_A\delta g_A + g_V\delta g_V$ and $g_V^\tau \approx 0.07 g_A^\tau$. The issue of a small universality breaking correction to $\Gamma_\tau$ will be a main topic of this paper.

It is often speculated that the third family, and the top quark in particular, is somehow wrapped up in the physics of electroweak symmetry breaking. In our extension the new third family flavor physics *is* tied up with electroweak symmetry breaking; the source of the mass of the $X$ boson is the same as for the $W$ and $Z$ masses. All these boson masses arise due to their axial couplings to fermions, and this gives us the relation [3]

$$\frac{g_X}{M_X} \;=\; \frac{e}{4csM_Z} \tag{2}$$

---

[2] There are no gauge anomalies in the model of [5]. The point is that the $X$ boson also couples to a fourth family, and the mass generation mechanism for the fourth family determines the remaining lighter third family fields and their $X$ charges.

where $c = \cos\theta_W$ and $s = \sin\theta_W$. With this relation the corrections induced by the $X$ boson depend only weakly on the unknown $X$ mass.

We are motivated in our study in the following ways.

- It seems that the standard model is far from confirmed by present data, since a new gauge boson with a coupling-to-mass ratio the same as the known weak bosons can still be added without conflict with the "precision data".
- An extension of the standard model with a new nonuniversal gauge interaction is completely at odds with the GUT/supersymmetry orthodoxy; this should provide incentive understand the effects of such an extension, if only to rule it out.
- The present precision data [6] actually favors this extension over the non-extended standard model.[3]

The main effects of the $X$ boson arise through the mixing with the $Z$, which causes a shift of the $Z$ couplings to the third family. The relative sizes of the various shifts are determined by the $X$ couplings and only the overall magnitude depends on the estimate of the $X$-$Z$ mixing in [3]. All the $X$ boson induced shifts quoted below can be simply scaled if a different estimate of the $X$-$Z$ mixing is used. Perhaps the result of most interest is $\delta\Gamma_b/\Gamma_b = 0.021$ to be compared to the experimental value of $0.020\pm0.009$. To obtain the latter we attribute any experimental deviation from the standard model value of $\Gamma_b/\Gamma_h$ to a shift in $\Gamma_b$. The increase in $\Gamma_b$ increases the predicted $\Gamma_h/\Gamma_\ell$ and thus decreases the $\alpha_s(M_Z)$ extracted from $\Gamma_h/\Gamma_\ell$ by an amount $\delta\alpha_s(M_Z) = -0.014$. $\Gamma_h$ is unaffected due to the canceling shifts in $\Gamma_b$ and $\alpha_s(M_Z)$, and as noted in [3], the result is that the extracted $\alpha_s(M_Z)$ is brought into line with low energy measurements. In this way a single new physics contribution to the $Zb\bar{b}$ vertex resolves the two features of the present data potentially embarrassing to the standard model [3][7]. Whether or not the $\alpha_s$ measurement is embarrassing is controversial, but see [8].

For the $\tau$ we find that $X$ boson substantially increases the asymmetry parameter, $\delta\mathcal{A}_\tau/\mathcal{A}_\tau = 0.21$. Here the jury is still out since the two independent measurements of $\mathcal{A}_\tau/\mathcal{A}_e$, from the forward-backward asymmetries (assuming $e$-$\mu$ universality) and from the tau polarization studies, give results not in good agreement; $\delta\mathcal{A}_\tau/\mathcal{A}_\tau = 0.57\pm0.26$ and $0.06\pm0.11$ respectively. For the $\tau$ neutrino we obtained $\delta\Gamma_{\nu\tau}/\Gamma_{\nu\tau} = -0.015$, which is not in conflict with $-0.014\pm0.023$. The latter is inferred from the measured ratio $\Gamma_{\text{inv}}/\Gamma_\ell$ assuming three light neutrinos. Note that the experimental values for all these universality breaking shifts are obtained from observables which are quite insensitive to possible oblique (universal) corrections. The $X$ boson also produces a slight shift in $\mathcal{A}_b$, but here there is no well-measured observable insensitive to oblique corrections.

---

[3] Of course the larger theory in which this extension is embedded may have additional effects. But note that the larger theory proposed in [5] is not a standard ETC theory.

For this and other observables it is necessary to know the oblique corrections, and this in turn requires a global fit. Without actually performing a global fit it is possible to see that the impact of an $X$ boson on other observables will be small. For example the shift in $\delta\mathcal{A}_b/\mathcal{A}_b = -0.0054$ from the $X$ boson can be compared to the present 4% error in the observable $A_{FB}^b$ ($\propto \mathcal{A}_b$). The shift in the leptonic widths (mostly $\Gamma_{\nu\tau}$) implies a $-0.09\%$ shift in the total width $\Gamma_Z$ to be compared with its present 0.15% error. We have seen that $\alpha_s(M_Z)$ should be allowed to vary in a global fit, which will imply that any shift in $\Gamma_h$ is minimal. Thus the total shift in $\Gamma_Z$, and the $\rho$ parameter extracted from $\Gamma_Z$, will be small. A global analysis has been performed in [9] which accounts for oblique as well as nonoblique corrections to $Zb\bar{b}$ couplings. The resulting constraints, not surprisingly, are completely compatible with the $X$ boson shifts $\delta g_L^b = -\delta g_R^b = -0.0038$ (here $g_{L,R}^b$ is defined as in [9]). Note that the global fit in [9] did not allow $\alpha_s(M_Z)$ to vary, which was instead fixed at 0.012.[4]

Lastly we may ask how the values of $\sin^2\theta_W$ extracted from various observables should be corrected to account for the existence of the $X$ boson. For the observable $A_{FB}^b = (3/4)\mathcal{A}_b\mathcal{A}_e$ most of the dependence on $\sin^2\theta_W$ comes from $\mathcal{A}_e$, and thus the presently extracted value of $\sin^2\theta_W$ needs only be reduced by 0.0001 to obtain the true value. The $\sin^2\theta_W$ extracted from $A_{FB}^\tau = (3/4)\mathcal{A}_\tau\mathcal{A}_e$ must be increased by 0.0019. This brings the $\sin^2\theta_W$ from $A_{FB}^\ell$ (average over three leptons) even closer to the $\sin^2\theta_W$ from $A_{FB}^b$. The $\sin^2\theta_W$ extracted from the average $\tau$ polarization $\mathcal{P}_\tau = -\mathcal{A}_\tau$ must be increased by 0.0038. This would put this $\sin^2\theta_W$ about 1.5$\sigma$ above the average of the other LEP measurements. We already noted that $\mathcal{A}_\tau/\mathcal{A}_e$ extracted from $\tau$ polarization studies does not support the $X$ boson hypothesis, and we now see that the source of the discrepancy is $\mathcal{P}_\tau$. It also appears that improvements in the measurement of the average $\tau$ polarization are beginning to be limited by systematic uncertainties [6].

We finally turn to the vertex corrections, diagrams in which the $X$ boson is attached directly to the third family fermions. These effects are generally smaller than the $X$-$Z$ mixing effects. In fact they occur at order $q^2$ in a momentum expansion, where $q$ is the momentum flowing into the vertex, whereas the $X$-$Z$ mixing effects occur at order $q^0$. On the other hand the vertex corrections affect all of the $Z$ couplings to the third family, and not just those couplings proportional to the $X$ boson couplings. Of particular interest then is the vertex correction to the axial $Z\tau\bar{\tau}$ coupling, which is not affected by $X$-$Z$ mixing. As we have said, the partial width data is very sensitive to any such correction. The $X$ induced vertex corrections are also potentially interesting in their effect on $t\bar{t}$ production in $p\bar{p}$ colliders. Here it is not obvious that the corrections to the $gt\bar{t}$ vertex are negligible, since the energies involved may be comparable to the $X$ mass, and the $X$ may be strongly interacting.

---

[4] We also do not understand the claim made in [7] that the anomaly in $\Gamma_b$ should be accounted for by a $\delta g_R^b$ and not a $\delta g_L^b$. Since $\Gamma_b$ is less sensitive to a $\delta g_R^b$, if $\delta g_L^b = 0$ then the required $\delta g_R^b$ would imply a 4% decrease in $\mathcal{A}_b$.

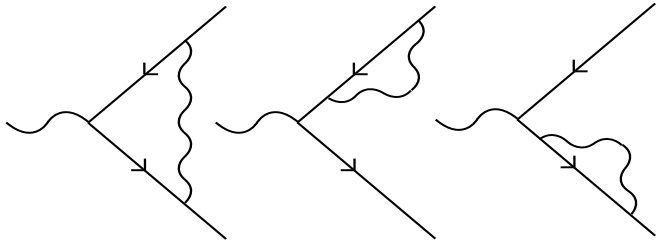

Figure 1 : Diagrams contributing to $Z$ or gluon vertex corrections induced by the $X$ boson. The fermion is a member of the third family.

We will start with the $Z$ decay case in which we can safely ignore the mass of the fermion ($b$, $\tau$, $\nu$) compared to the $Z$ and $X$ masses. Here we can obtain analytical expressions for the corrections both before and after expanding in $M_Z/M_X$. We consider the diagrams in Fig. (1) and write the massive $X$ propagator in unitary gauge. We express the integrals in terms of scalar integrals and find that the individual diagrams are finite. The self-energy graphs contribute a finite field renormalization which is accounted for in the usual way. In this case of ignoring the fermion mass it makes no difference whether we consider vector or axial-vector couplings for the $X$ and $Z$. For $Z$ decay into any of the third family fermions these corrections increase each partial width by the same amount, while the asymmetries are unaffected. For the partial widths

$$\frac{\delta\Gamma}{\Gamma} = \frac{g_X^2}{16\pi^2}\{-4\left(1 + \tfrac{1}{r}\right)^2[\ln(1 + r)\ln(r) + \mathrm{dilog}(1 + r)] \\ + 2\left(3 + \tfrac{2}{r}\right)\ln(r) - 7 - \tfrac{4}{r}\,\} \tag{3}$$

where $r \equiv (M_Z/M_X)^2$ and the dilog function is defined by

$$\mathrm{dilog}(x) = \int_1^x \frac{\ln(y)}{1-y}\,dy. \tag{4}$$

If we expand this result in powers of $r$ we find

$$\frac{\delta \Gamma}{\Gamma} = \frac{g_X^2}{72\pi^2} r(11 - 6\ln(r)). \tag{5}$$

Note the term nonleading in $\ln(r)$ is significant. If we use (2) to determine the value of $g_X$ then the linear dependence on $r$ is canceled. For example with $M_X/M_Z = 5$ we obtain $\delta\Gamma/\Gamma = 0.0015$.

This correction to the partial width must be added to the correction coming from $X$-$Z$ mixing. For both $\Gamma_b$ and $\Gamma_{\nu\tau}$ the mixing effect is about 10 times as large. But for $\Gamma_\tau$ the $X$-$Z$ mixing only affects the vector $Z\tau\bar\tau$ coupling which gives $\delta\Gamma_\tau/\Gamma_\tau = 0.0022$ [3], and thus the vertex correction is of comparable importance. This is the case even though the vertex correction is formally suppressed by $(M_Z/M_X)^2$. Our total correction to $\delta\Gamma_\tau/\Gamma_\tau$ is in the 0.003 to 0.004 range. The precision of the current measurements is starting to become comparable; the current data gives $\delta\Gamma_\tau/\Gamma_\tau \approx 0.004 \pm 0.004$.

It is amusing to note that the significance of the data used in determining $R \equiv \Gamma_h/\Gamma_\ell$ has been recently called into question [10]. The distribution of the individual values of $R_e$, $R_\mu$, $R_\tau$ from each of the four experiments were found to be inconsistent with gaussian statistics. The conclusion of [10] was that there must be substantial systematic effects unaccounted for in the original error estimate. An alternative conclusion, if there is indeed a problem, is that the interpretation of the data is wrong; if the assumption of universality is relaxed then the apparent problem with the data largely disappears.

We now turn to the case of top production. For nonzero quark mass the self-energy graphs are no longer finite, and we renormalize by requiring that the correction to the gluon vertex vanish at $s \equiv q^2 = 0$. Although in our model the $X$ boson couplings to the quarks are axial, we will also consider vector $X$ couplings for comparison. We have to resort to evaluating integrals numerically, and we present the results in Fig. (2). We plot the correction as a function of $t \equiv s/M_X^2$, and to determine $g_X$ we use (2) with $M_X/M_Z = 5$. The different lines represent different choices of $u \equiv 4m^2/s < 1$ where $m$ is the quark mass. We also display the $m = 0$ case, which is given by (3) with $r$ replaced with $t$. Thus we find that quark mass effects give much more enhancement for vector rather than axial $X$ couplings. In fact in the axial case for $u = 0.8$ the result is very similar to $m = 0$; for decreasing $u$ the result decreases further (but not much smaller than the $u = 0.7$ line) until it again approaches the $m = 0$ line for small $u$. Our conclusion is that the vertex corrections induced by the $X$ boson with axial quark coupling make only a very minor correction to $t\bar t$ production.

In contrast we have found that the vertex corrections contribute to a potentially observable shift in the $Z$ partial width to $\tau$. Our results for the vertex corrections could be easily adapted to cases other than the specific $X$ boson we have chosen to study. For example if the coupling of a new gauge boson was not constrained by (2), then the vertex corrections considered here could be much more significant. As for our specific $X$ boson and the $X$-$Z$ mixing effects, it will be interesting to keep an eye on certain trends in the data as the data improves.

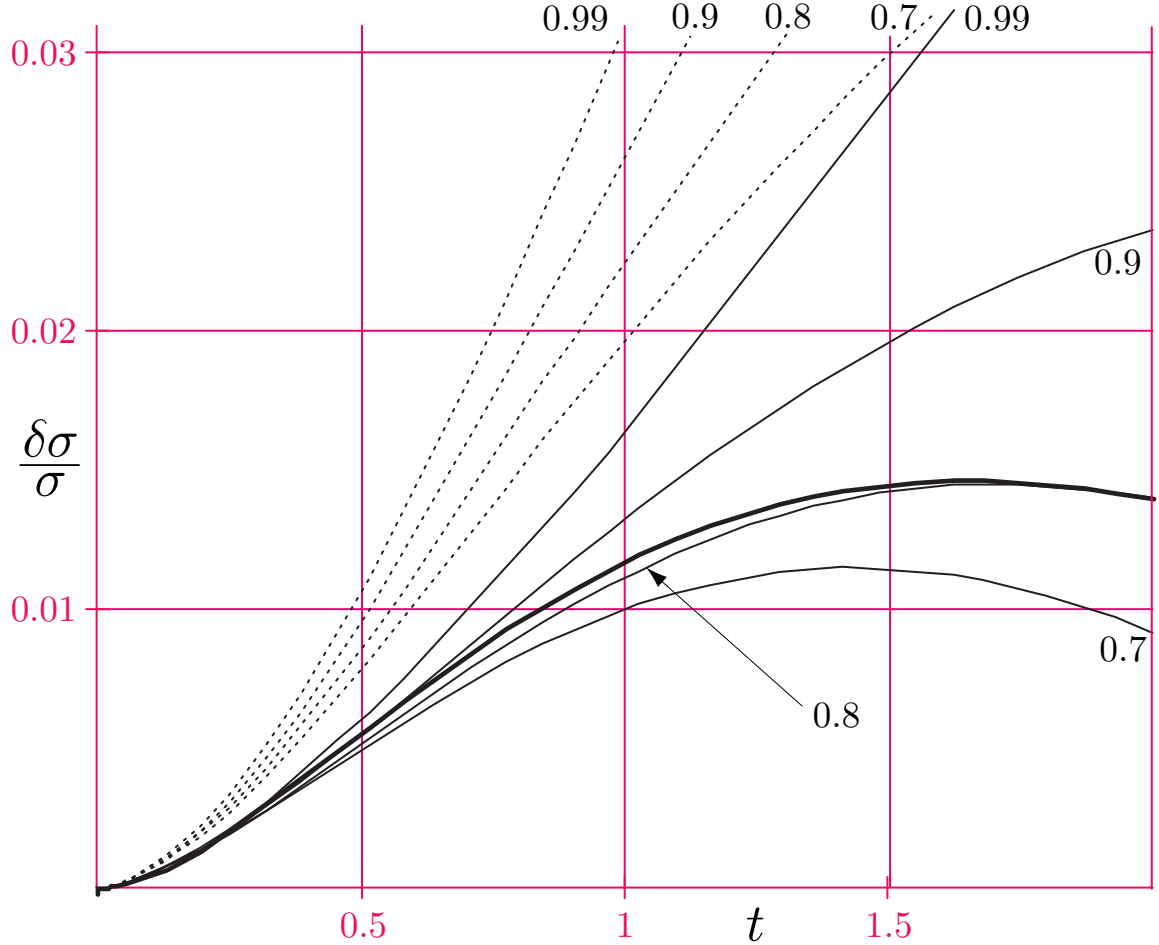

Figure 2 : $t \equiv s/M_X^2$ and the curves are labelled by the value of $u \equiv 4m^2/s$. The heavy solid line corresponds to $m = 0$. The $X$ boson with axial coupling to the top quark (thin solid lines) is to be compared with the vector coupling case (dotted lines).

# Acknowledgements

I thank John Terning for describing to me his global fit to the data. This research was supported in part by the Natural Sciences and Engineering Research Council of Canada.